\documentclass[floatfix,aps,prl,superscriptaddress,twocolumn,showpacs]{revtex4-1}

\usepackage[none]{hyphenat}
\usepackage{graphicx}
\usepackage{dcolumn}
\usepackage{bm}
\usepackage{amsmath}
\usepackage{color}
\usepackage[colorlinks,
linkcolor=blue,
citecolor=blue,
urlcolor=blue]{hyperref}
\makeatother

\begin{document}
\title{Branching of high-current relativistic electron beam in porous materials}

\author{K. Jiang}
\affiliation{Shenzhen Key Laboratory of Ultraintense Laser and Advanced Material Technology, Center for Advanced Material Diagnostic Technology, and College of Engineering Physics, Shenzhen Technology University, Shenzhen 518118, People's Republic of China}
\affiliation{College of Applied Sciences, Shenzhen University, Shenzhen 518060, People's Republic of China}

\author{T. W. Huang}
\email{taiwu.huang@sztu.edu.cn}
\affiliation{Shenzhen Key Laboratory of Ultraintense Laser and Advanced Material Technology, Center for Advanced Material Diagnostic Technology, and College of Engineering Physics, Shenzhen Technology University, Shenzhen 518118, People's Republic of China}

\author{R. Li}
\affiliation{Shenzhen Key Laboratory of Ultraintense Laser and Advanced Material Technology, Center for Advanced Material Diagnostic Technology, and College of Engineering Physics, Shenzhen Technology University, Shenzhen 518118, People's Republic of China}

\author{M. Y. Yu}
\affiliation{Shenzhen Key Laboratory of Ultraintense Laser and Advanced Material Technology, Center for Advanced Material Diagnostic Technology, and College of Engineering Physics, Shenzhen Technology University, Shenzhen 518118, People's Republic of China}

\author{H. B. Zhuo}
\affiliation{Shenzhen Key Laboratory of Ultraintense Laser and Advanced Material Technology, Center for Advanced Material Diagnostic Technology, and College of Engineering Physics, Shenzhen Technology University, Shenzhen 518118, People's Republic of China}

\author{S. Z. Wu}
\affiliation{Shenzhen Key Laboratory of Ultraintense Laser and Advanced Material Technology, Center for Advanced Material Diagnostic Technology, and College of Engineering Physics, Shenzhen Technology University, Shenzhen 518118, People's Republic of China}

\author{C. T. Zhou}
\email{zcangtao@sztu.edu.cn}
\affiliation{Shenzhen Key Laboratory of Ultraintense Laser and Advanced Material Technology, Center for Advanced Material Diagnostic Technology, and College of Engineering Physics, Shenzhen Technology University, Shenzhen 518118, People's Republic of China}
\affiliation{College of Applied Sciences, Shenzhen University, Shenzhen 518060, People's Republic of China}

\author{S. C. Ruan}
\email{scruan@sztu.edu.cn}
\affiliation{Shenzhen Key Laboratory of Ultraintense Laser and Advanced Material Technology, Center for Advanced Material Diagnostic Technology, and College of Engineering Physics, Shenzhen Technology University, Shenzhen 518118, People's Republic of China}
\affiliation{College of Applied Sciences, Shenzhen University, Shenzhen 518060, People's Republic of China}

\date{\today}

\begin{abstract}
Propagation of high-current relativistic electron beam (REB) in plasma is relevant to many high-energy astrophysical phenomena as well as applications based on high-intensity lasers and charged-particle beams. Here we report a new regime of beam-plasma interaction arising from REB propagation in medium with fine structures. In this regime, the REB cascades into thin branches with local density hundred times the initial value and deposits its energy two orders of magnitude more efficiently than that in homogeneous plasma, where REB branching does not occur, of similar average density. Such beam branching can be attributed to successive weak scatterings of the beam electrons by the unevenly distributed magnetic fields induced by the local return currents in the skeletons of the porous medium. Results from a model for the excitation conditions and location of the first branching point with respect to the medium and beam parameters agree well with that from pore-resolved particle-in-cell simulations.
\end{abstract}

\maketitle
\sloppy{}

Transport of high-current relativistic electron beam (REB) in plasma is a fundamental issue in high-energy-density space and laboratory plasmas \cite{Bret}, and has attracted much research interest in areas such as solar flare physics \cite{Karlicky}, astrophysics \cite{Marti}, and quantum chromodynamics \cite{Mrowczynski,Mannarelli}. The problems involved include collisionless shocks \cite{Nishikawa,Li}, cosmic magnetic field generation \cite{Lazar}, gamma-ray bursts \cite{Silva,Lyubarsky}, etc. It is also relevant to many current applications, including inertial confinement fusion \cite{Robinson} and compact particle and/or radiation sources \cite{Golde, Benedetti}.

The dynamics of REB propagation in matter and the resulting energy transfer have been a long-standing research topic \cite{Jagher}. Nonlinear processes can lead to kinetic-scale plasma instabilities that modulate the beam and electromagnetic fields, resulting in energy/momentum transfer from the beam to the medium \cite{Bret,Weibel,Watson,Davidson,Gremillet,Gremillet1,Krasheninnikov,Bret1,Ran,Gong}. Experiments on these processes show that, depending on the material traversed by the REB, distinct beam behaviors can appear. In particular, beam breakup occurs more likely in foams than in metals \cite{Jung,Fuchs,Manclossi,Romagnani,Ruyer}. For dense foams, since the foam's structure is usually smaller than the spatial scale of interest, they are often treated as a continuum of homogeneous density (or with a gradient) in theoretical analyses. That is, the foam’s possible fine structuring is not taken into account. 

Low-density foams are often used in experiments as subcritical-density materials \cite{Bin, Bin1,Rosmej,Rosmej1}, which however consist of sponge-like sub-micron sized intertwining solid-density skeletons with micron sized empty regions (hereafter referred to as pores) in between \cite{Nagai}. Previous works suggest that structures at microscale can significantly affect macroscale phenomena \cite{YTLi,Robinson1,Chatterjee,Kemp,Belyaev,Cipriani}. Thus, low-density foams can no longer be treated as a continuum, especially when the pore size is considerably larger than the skin depth of typical laser-produced electron beams. In this case, the self-generated fields during the REB propagation can be affected by the microscopic structure, resulting in randomly uneven but long-range correlated field distributions that can lead to complex/unusual transport behavior \cite{Heller}.

In this Letter, we report a new regime of beam-plasma interaction arising from the microscopic internal structure of porous materials. As a high-current REB propagates in such a medium, thin branched flows with local density hundred times that of the pristine beam at the bifurcation points, or caustics, can form. The beam energy is efficiently transferred to that of the ionized-skeleton particles and the self-generated electric and magnetic fields. The energy conversion efficiency can be two orders of magnitude greater than that in a homogeneous medium of similar average density but without the skeleton-and-pore heterogeneity. Our particle-in-cell (PIC) simulation results agree well with that from a simple model, where the REB branching is attributed to scattering of the beam electrons by the self-generated uneven fields in the pores.

\begin{figure*}
\centering
\includegraphics[width=17.2cm]{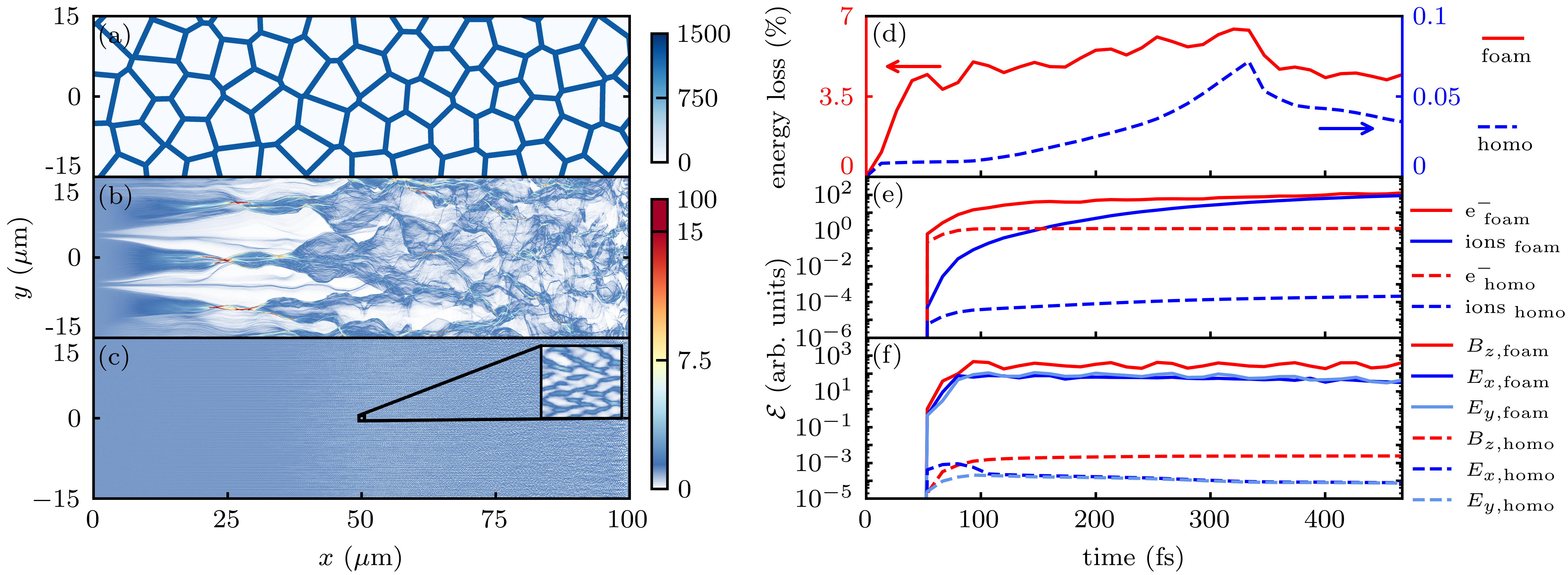}
\caption{(a) Density (in units of $n_{b0}$, same below) of Si atoms in the pristine foam. Densities at $t=333$ fs of the REB as it propagates in (b) foam and (c) a homogeneous target at the same average density of the foam. The inset in (c) shows an enlargement of the marked region ($49.5\ \mu$m $<x<50.5\ \mu$m, $-0.5\ \mu$m $<y<0.5\ \mu$m). The electron density at the caustics in (b) is up to $\sim 100n_{b0}$. (d) Evolution of the energy loss $\eta = 1-\langle E_{k,b}\rangle / \langle E_{k,b0}\rangle$ of the REB. Here $\langle E_{k,b}\rangle$ is the average kinetic energy of all the beam electrons in the simulation box and $\langle E_{k,b0}\rangle$ corresponds to its initial value. (e) Evolution of the target electron and ion kinetic energies $\mathcal{E}_{\mathrm{e}^{-}}$ and $\mathcal{E}_{\mathrm{ions}}$ in the region $13.5\ \mu$m $<x<25.5\ \mu$m, $-5.5\ \mu$m $<y<5.5\ \mu$m. (f) Evolution of the azimuthal magnetic energy $\mathcal{E}_{B_{z}}$, longitudinal and lateral electric energy $\mathcal{E}_{E_{x}}$ and $\mathcal{E}_{E_{y}}$ in the region.} \label{fig_1}
\end{figure*}

To study the transport of high-current REB in porous media, we first conduct pore-resolved two-dimensional (2D) PIC simulations using \textsc{epoch} \cite{Arber}. In the simulation, the porous material is SiO$_{2}$ foam, modeled by random Voronoi cells \cite{Song} with average pore size of $l_{c}=8\ \mu$m, as shown in Fig. \ref{fig_1}(a). The number densities of Si and O atoms in the $1\ \mu$m-thick skeletons are $n_{\mathrm{Si}}=2.15\times10^{28}$ m$^{-3}$ and $n_{\mathrm{O}}=4.3\times10^{28}$ m$^{-3}$, respectively, corresponding to porosity up to $90\%$. The pristine REB is monoenergetic and of uniform density $n_{b0}=1.72\times10^{25}$ m$^{-3}$ and momentum $p_{x0}=\gamma m_ec$, where the Lorentz factor is $\gamma=100$, $m_e$ is the electron rest mass, and $c$ is the vacuum light speed. The beam duration is 400 fs. Such beams can be potentially generated in laboratory from linear accelerators or laser acceleration \cite{note,Huang}. The simulation box is of size 0 $\mu$m $<x<$ 100 $\mu$m and -15 $\mu$m $<y<$ 15 $\mu$m, with $8000\times2400$ grid cells, which resolves the skin-depth of the solid skeletons. There are 14 macroparticles per cell for both Si and O atoms, and 100 for the beam electrons. Periodic boundary conditions in the lateral direction are used. Field ionization, included in \textsc{epoch} \cite{Arber}, converts the foam into plasma. The collisional effects \cite{Perez} do not influence the REB propagation here (see Appendix A and Supplemental Material \cite{Sup}).

Figure \ref{fig_1}(b) for the density distribution of the beam electrons at $t=333$ fs shows a branched flow pattern \cite{Heller,Topinka,Barkhofen,Deguelder,Derr,Patsyk,Patsyk1,Jiang}: the REB breaks up into three narrow branches at $x\sim 23\ \mu$m. At the caustics, the highly focused beam electrons can have density up to \mbox{$\sim100n_{b0}$}. It is of interest to note that instead of merging, the beam branches propagate almost independently even as they broaden and cross each other, eventually reaching a random pattern downstream at $x>50\ \mu$m. That is, the evolution of the branches appears to be linear. It is clarified that no beam-plasma instabilities are observed during the REB branchings, as the presence of skeleton-and-pore heterogeneity limits
the free evolution of the plasma and induced fields \cite{Sup}. Simulations using different samples of the foam structure (but of the same statistic properties, e.g., average density and pore size) indicate that the average lateral separation between two adjacent caustics is about $9\ \mu$m, which matches the pore size of the foam. That is, the beam branching is indeed related to the microscopic foam structure, and is also in agreement with the statistical property of branched flows in other systems \cite{Heller,Kaplan,Metzger,Metzger1}. This regime is quite robust with respect to the beam parameters \cite{Sup}. Moreover, Fig. \ref{fig_1}(c) for REB propagation in homogeneous plasma with the same average density as the foam shows that branched flow is absent, and the REB suffers small-scale fluctuations as expected \cite{Watson,Gremillet1}.

The REB can also efficiently deposit its energy in foam targets. Figure \ref{fig_1}(d) shows that the energy loss $\eta$ of the REB in the foam target is two orders of magnitude greater than in homogeneous ones. The decrease of $\eta$ for $t>333$ fs is due to electrons leaving the simulation box from the right boundary. By probing a selected region in the target, one sees from Fig. \ref{fig_1}(e) that as the REB enters the region at $t\sim 45$ fs, rapid ionization of the skeletons results in a jump in the foam-electron energy $\mathcal{E}_{\mathrm{e}^{-}}$, followed by further increase to a high level. The foam ions are energized on a longer time scale, with their energy $\mathcal{E}_{\mathrm{ions}}$ approaching $\mathcal{E}_{\mathrm{e}^{-}}$ in sub-picoseconds. $\mathcal{E}_{\mathrm{e}^{-}}$ is about hundred times, and $\mathcal{E}_{\mathrm{ions}}$ up to six orders of magnitude more than, that in homogeneous plasma. Moreover, since the skeleton-and-pore structure allows for highly localized strong fields (to be discussed in detail later), the energy of the generated magnetic field is even larger than that of the foam particles, whereas in homogeneous plasma it is hardly observable, as shown in Fig. \ref{fig_1}(f).

\begin{figure}
\centering
\includegraphics[width=8.6cm]{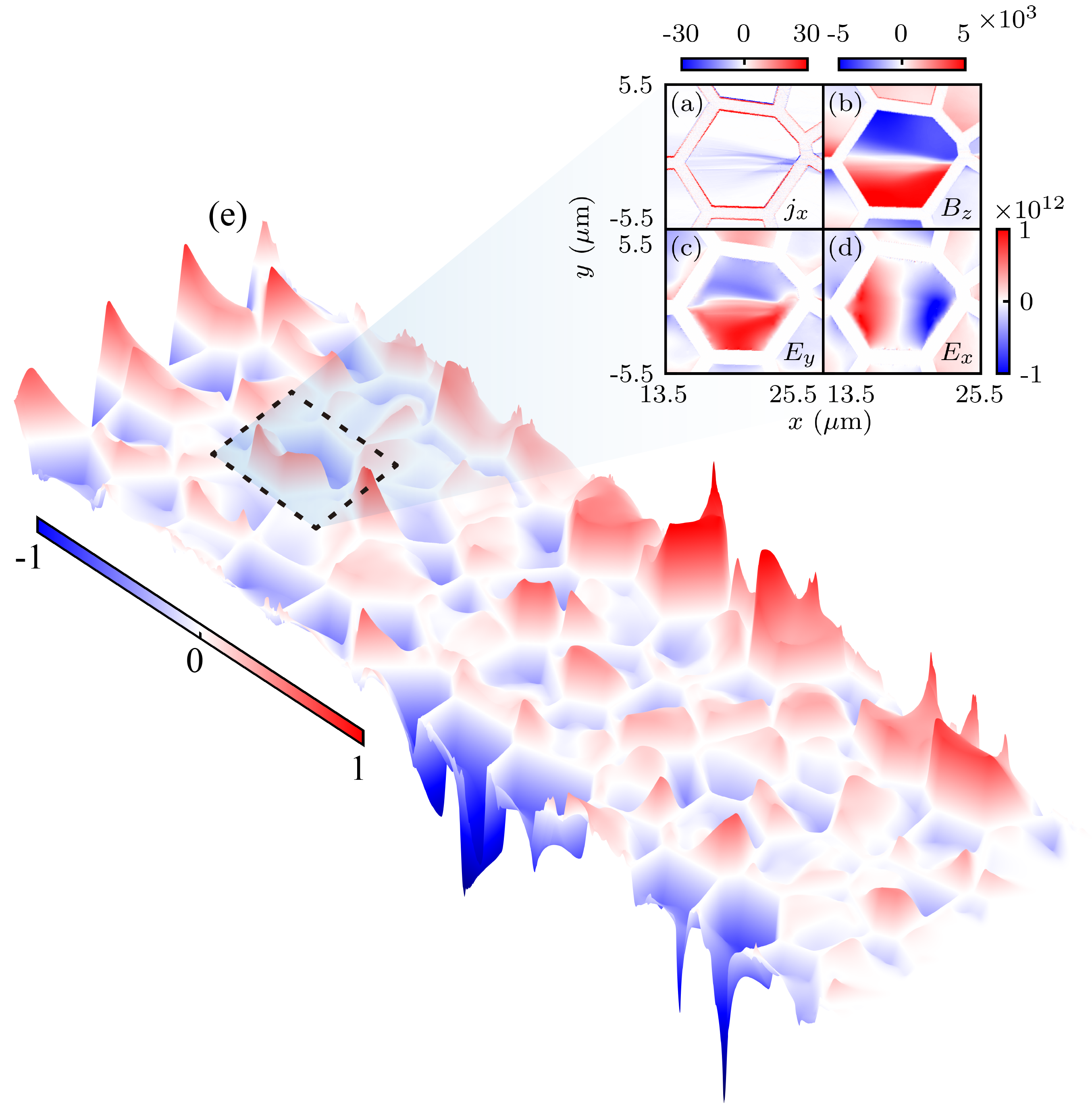}
\caption{Distributions of (a) longitudinal current density $j_{x}$ (in units of $|j_{b0}|$), (b) azimuthal magnetic field $B_{z}$ (in Tesla), (c) lateral and (d) longitudinal electric field $E_{y}$ and $E_{x}$ in (V/m) at $t=333$ fs in the selected region in (e). (e) The lateral force $F_{y}=e(cB_{z}-E_{y})$ (in arbitrary units) acting on the beam electrons. The dashed lines mark the selected region ($13.5\ \mu$m $<x<25.5\ \mu$m, $-5.5\ \mu$m $<y<5.5\ \mu$m).} \label{fig_2}
\end{figure}

The mechanism for onset of the branching is two-fold: (1) skeleton ionization by the incident REB and generation of electric and magnetic fields, and (2) scattering of the beam electrons by the latter. As the REB impinges on the foam and ionizes the affected skeletons into solid-density plasma, cold return current localized in the skin layer of the skeleton is induced to compensate the beam current that far exceeds the Alfv\'{e}n limit \cite{Alfven}. Figure \ref{fig_2}(a) shows that the return current is of huge density, namely  $j_{r}\sim|j_{b0}|l_{c}/2l_{s}\exp(1)\sim 105|j_{b0}|$, where $j_{b0}=-\sqrt{\gamma^{2}-1}n_{b0}ec/\gamma\sim -8.3\times10^{14}$ A/m$^{2}$ is the beam-current density, $l_{s}=c/\omega_{pe}\sim 0.014\ \mu$m, $\omega_{pe}=\sqrt{n_{p}e^{2}/m_e\varepsilon_{0}}$ and \mbox{$n_{p}\sim 1.4\times10^{29}$ m$^{-3}$} are the skin length, plasma frequency and electron density of the skeletons, $-e$ is the electron charge, and $\varepsilon_{0}$ is the vacuum permittivity. Figure \ref{fig_2}(b) shows that the current generates strong azimuthal magnetic field in the pores. From Amp\'{e}re's law, one can estimate the peak magnetic strength as $|B_{z}|\sim\mu_{0}j_{b0}l_{c}/2\sim 4.1\times10^{3}\ \mathrm{T}$, in agreement with the simulation results. The REB also provides net charge in the pores for electrostatic fields of amplitude $|E_{x}|\text{\ensuremath{\sim |E_{y}|}}\sim en_{b0}l_{c}/4\varepsilon_{0}\sim 6.2\times10^{11}$ V/m as given by Poisson's equation, in agreement with Figs. \ref{fig_2}(c) and (d). Thus, the lateral force acting on the beam electrons can be attributed to the magnetic field associated with the return current, namely, $|F_{y}|=e(c|B_{z}|-|E_{y}|)\sim  ec|B_{z}|/2$. Since the self-generated fields are distributed in the foam's randomly distributed pores, the distribution of $F_{y}$ has a randomly uneven pattern that is long-range correlated, as shown in Fig. \ref{fig_2}(e).

\begin{figure}
\centering
\includegraphics[width=8.6cm]{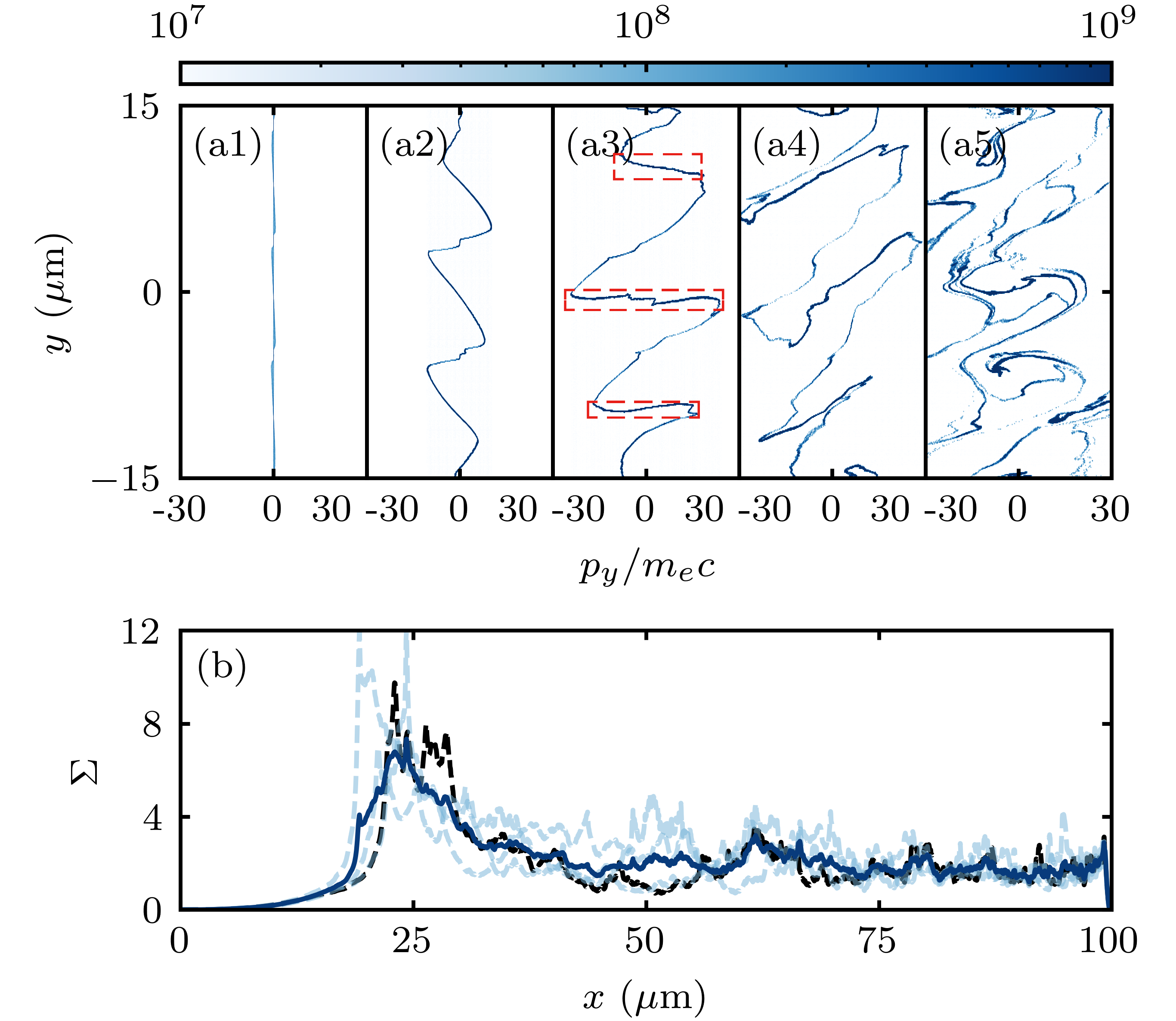}
\caption{Phase space $(y,p_{y})$ of the beam electrons at (a1) $x=0.1\ \mu$m, (a2) $4.3\ \mu$m, (a3) $24.3\ \mu$m, where the red dashed lines mark the singularity regions, (a4) $44.3\ \mu$m, and (a5) $80\ \mu$m. (b) Scintillation index $\Sigma$ corresponding to the beam electron density (black dashed curve). The light-blue dashed curves are results from three other foam samples with the same statistic properties. The blue solid curve shows the ensemble average of the four samples.} \label{fig_3}
\end{figure}

We now consider the random scattering of the beam electrons by the uneven fields discussed above, i.e., the process (2). REB branching can be seen from the phase space $(y,p_{by})$ of the beam electrons. As shown in Fig. \ref{fig_3}(a1), the initial $(y,p_{by})$ is nearly a straight line, since the beam is initially monoenergetic. Figure \ref{fig_3}(a2) shows that as the beam propagates, successive random kicks from the lateral force $F_{y}$ stretch and curve their phase-space distribution. Figure \ref{fig_3}(a3) shows that at $x\sim 23\ \mu$m three singularities, where $\partial_y p_{by}\rightarrow\infty$, occur. The phase-space singularities correspond to caustics where the REB bifurcates, as shown above in Fig. \ref{fig_1}(b). It also agrees with Fig. \ref{fig_3}(b) for the spatial evolution of the scintillation index $\Sigma=(\langle n_{b}^{2}\rangle/\langle n_{b}\rangle^{2})-1\;$, which measures the relative strength of the beam density variation \cite{Green}. Figures \ref{fig_3}(a4) and (a5) show that the phase space distribution becomes random at $x>44.3\ \mu$m due to further random kicks from the fields as well as overlapping of different branches. Correspondingly, the energy spectrum of REB is greatly broadened \cite{comment1}.

\begin{figure}
	\centering
	\includegraphics[width=8.6cm]{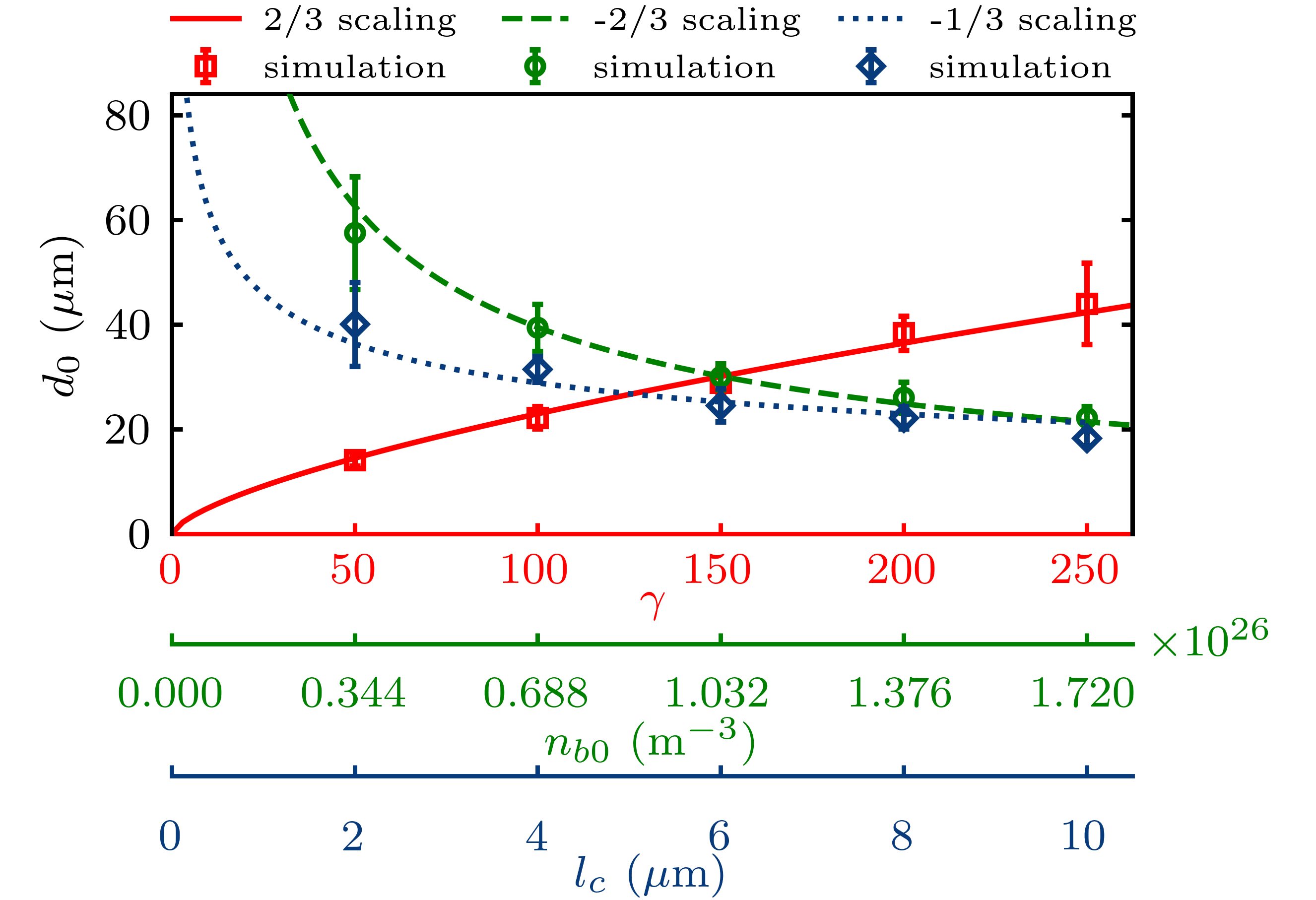}
	\caption{The distance $d_{0}$ for different beam Lorentz factors $\gamma$ (in red, with $n_{b0}=1.72\times10^{25}$ m$^{-3}$ and $l_{c}=8$ $\mu$m), initial beam density $n_{b0}$ (in green, with $\gamma=100$ and $l_{c}=8$ $\mu$m), and average pore size $l_{c}$ (in blue, with $n_{b0}=1.72\times10^{25}$ m$^{-3}$ and $\gamma=100$). The error bars and central markers represent the standard deviation and mean value of $d_{0}$, obtained from simulations of four foam samples. The curves are from Eq. \eqref{eq5}.} \label{fig_4}
\end{figure}

The onset of REB branching can be characterized by the distance $d_{0}$ from the foam's front surface to the first caustics. Here we are mainly interested in the lateral direction where branching occurs. From the relativistic momentum equation of collisionless electron fluid, one has

\begin{equation}
	\frac{\partial p_{by}}{\partial t}+v_{bx}\frac{\partial p_{by}}{\partial x}+v_{by}\frac{\partial p_{by}}{\partial y}=-\frac{\partial V_{y}}{\partial y},\label{eq1}
\end{equation}
where $v_{b}$ and $p_{b}$ are the REB velocity and momentum, and $V_{y}=-\int F_{y}\mathrm{dy}$ denotes a pseudopotential. Thermal pressure is negligible except perhaps at the caustics. For simplicity, we set $m_e=c=e=1$. Equation \eqref{eq1} can be decoupled from the continuity and Maxwell's equations since $V_y$ is slow varying compared with beam dynamics for $n_{b0} \ll n_p$.

For ultrarelativistic electron beams and weak pseudopotential $V_{y}$, namely $\gamma\gg1$, $p_{bx}\gg p_{by}$, and $v_{bx}\sim1\gg v_{by}$, the REB phase space curvature $u=\partial_yp_{by}$ is given by \cite{Sup}
\begin{equation}
\frac{\mathrm{d}u}{\mathrm{d}t}+\frac{1}{\gamma}u^{2}+\frac{\partial^{2}V_{y}}{\partial y^{2}}=0,\label{eq3}
\end{equation}
which is the Langevin equation with the stochastic pseudopotential $V_{y}$. Thus, the first caustics, where \mbox{$u\rightarrow\infty$}, occur at
\begin{equation}
d_{0}\propto l_{c}\left(\gamma/\overline{V}\right)^{2/3},\label{eq4}
\end{equation}
which is the well-known scaling law for branched flows \cite{Kaplan,Metzger,Metzger1,Mattheakis}. Substituting $F_y$ into $V_y$, one obtains $\overline{V}\equiv\sqrt{\left\langle V_{y}^{2}\right\rangle }\sim\mu_{0}e^{2}c^{2}l_{c}^{2}n_{b0}/8\sqrt{3}$, where $\left\langle \cdots\right\rangle$ denotes an average taken over the entire foam region. Thus, we have
\begin{equation}
d_{0}\propto l_{c}^{-1/3}n_{b0}^{-2/3}\gamma^{2/3}.\label{eq5}
\end{equation}

We see that $d_{0}$ decreases with increasing $l_c$ and $n_{b0}$, and increases with $\gamma$. When $l_{c}\rightarrow0$, namely the foam becomes rather dense, $d_{0}$ approaches infinity, indicating that beam branching is absent and beam-plasma instabilities would dominate in continuum medium. Although the above derivation assumes that the REB is initially monoenergetic, simulations show that Eq. \eqref{eq5} can also be applied to REBs with finite temperatures \cite{Sup}. In addition, Eq. \eqref{eq5} reveals several distinct features of REB branching, where the distance $d_{0}$ is independent of the skeleton density or beam temperature. The exponents in Eq. \eqref{eq5} also differ from existing beam-plasma instabilities \cite{Bret,Bret1,Bret2,Hao}. This suggests that REB branching is a new regime of relativistic beam-plasma interaction. Furthermore, our theoretical analysis imposes no restriction on the charged-beam species nor the profile of the pseudopotential, suggesting that branching could be common for high-energy charged-particle beam propagation in porous and similar materials.

To check the reliability of Eq. \eqref{eq5}, we have conducted over sixty simulations with different beam-target parameters. As shown in Fig. \ref{fig_4}, the results are in agreement with Eq. \eqref{eq5}. That is, once the parameters of the beam and porous material are given, the location of beam branching can be predicted.

From the above analysis, the requirement for REB branching can also be obtained. To prevent beam particle backscattering or localization on a scale of order $l_{c}$, $\overline{V}$ should be weak relative to $\langle E_{k,b0}\rangle$ \cite{Kaplan}, say $\overline{V}\lesssim 0.1\langle E_{k,b0}\rangle$. Thus, a prerequisite for REB branching is
\begin{equation}
	\gamma\gtrsim\mu_{0}e^{2}l_{c}^{2}n_{b0}/m_e. \label{eq6}
\end{equation}
For the simulation parameters in Fig. \ref{fig_1}, Eq. \eqref{eq6} gives \mbox{$\gamma\gtrsim39$}. Indeed, additional simulations show that REBs with too-low $\gamma$ can only propagate a distance $\text{\ensuremath{\sim l_c}}$ into the foam, as further penetration is inhibited by the strong self-generated fields. In addition, to maintain the foam structure during the interaction, the condition $n_{b0} \ll n_p$ is also required.

Since REBs with parameters satisfying Eq. \eqref{eq6} can be readily produced by current laser-plasma acceleration or linear accelerators, experimental detection of REB branching is possible. For example, one can irradiate low-density porous foam targets with REBs and measure the spatial distribution of the transition radiation at the foam’s rear surface. By adjusting the beam and foam parameters (e.g., beam energy and density, foam porosity and thickness), one can expect that the measured spatial distribution of the radiation strength resembles the properties of branched beam electrons.

\begin{figure}
	\centering
	\includegraphics[width=8.6cm]{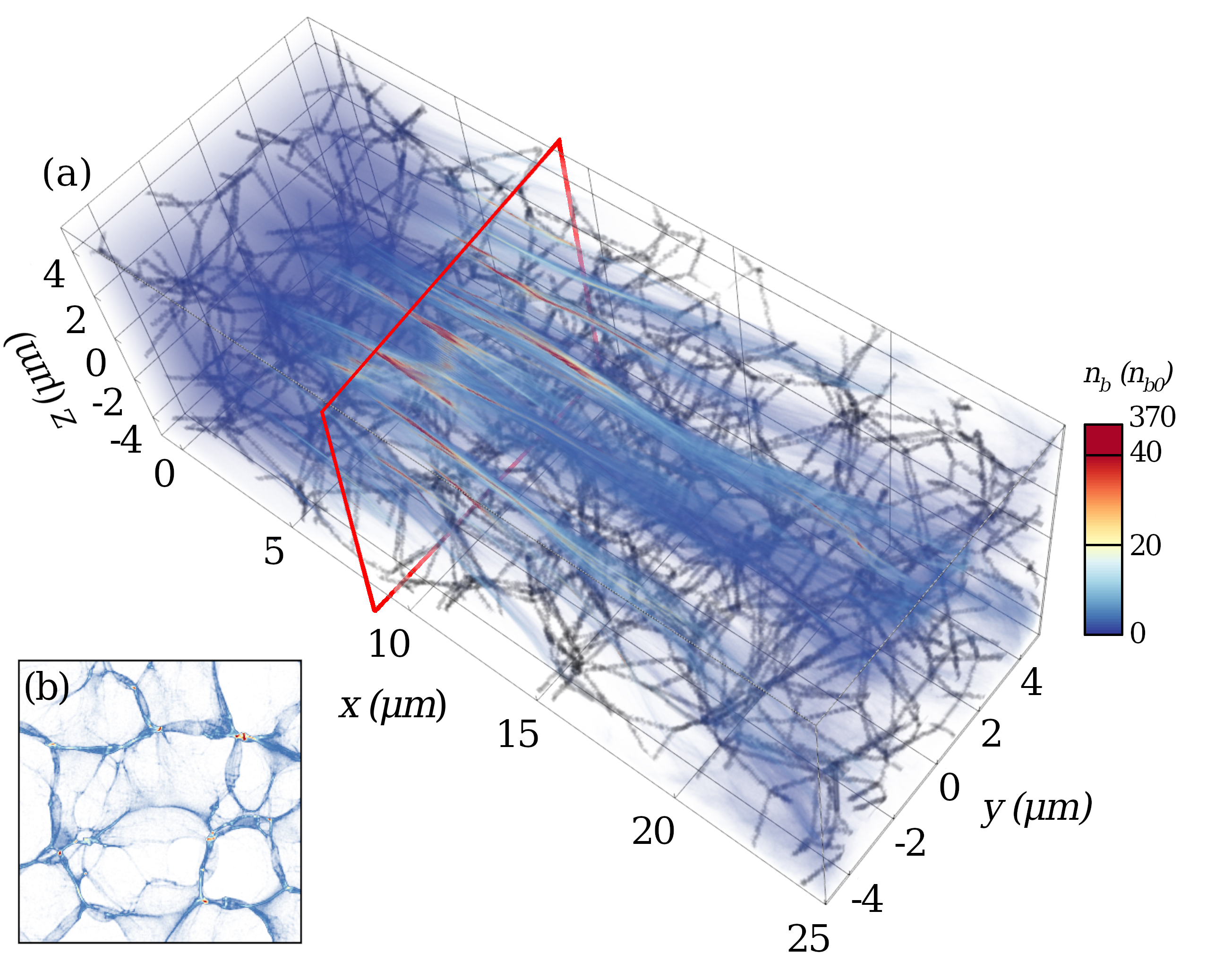}
	\caption{(a) Beam density (colored) at $t=120$ fs from a 3D simulation for REB branching in porous foam. The background gray color represents the pristine foam structure. (b) Transverse cut of the beam density at $x\sim8.7\ \mu$m (marked by the red rectangle) showing the caustics. The electron density at the caustics exceeds \mbox{$\sim370n_{b0}$}.} \label{fig_5}
\end{figure}

Figure \ref{fig_5} is for REB branching in 3D. Here an REB with $n_{b0}=3.4 \times 10^{26}$ m$^{-3}$ and $\gamma=5$ is used. The foam is modeled by a 3D Voronoi structure \cite{Gostick}. One sees that the results are similar to that in Fig. \ref{fig_1}(b) for the 2D case. Moreover, at the caustic surfaces shown in Fig. \ref{fig_5}(b), the beam density can be up to $\sim 370n_{b0}$. However, REB propagation in 3D porous materials involves more degrees of freedom than in 2D, so that Eq. \eqref{eq6} needs to be modified, which shall be discussed in a future work.

In conclusion, we have shown that high-current REB can efficiently deposit its energy in porous medium by forming thin and extremely-dense branches, thereby bringing the branched flow phenomenon to high-energy-density beam physics. REB branching is distinct from beam-plasma instabilities in terms of the beam density pattern, onset mechanism and parameter spaces, as well as scaling laws (see Appendix B). This interaction regime may open new prospects of charged-particle beam interaction with fine-structured materials, which should be of interest to inertial confinement fusion research, generation of bright broadband radiation, and laboratory modeling cosmic ray propagation in interstellar dust clouds, as well as raise attention to the physics of branched flows, which can be considered as a regime of transport between ballistic and diffusive, and has only recently been identified as such \cite{Heller}.

This work is supported by the National Natural Science Foundation of China (Grants No. 12175154, No. 11875092, No. 12005149, and No. 12205201), and the Natural Science Foundation of Top Talent of SZTU (Grant No. 2019010801001 and 2019020801001). The \textsc{epoch} code is used under UK EPSRC contract (EP/G055165/1 and EP/G056803/1). K. J. would like to thank A. Pukhov, X. F. Shen, L. Reichwein at the Heinrich-Heine-Universit\"{a}t D\"{u}sseldorf, and H. Peng at SZTU for useful discussions.

\vspace{1em}

\textit{Appendix A: Collisional effects.\textemdash }The mean free path for the beam electrons reads
\begin{equation}
	\lambda_{\mathrm{mfp},b} = \frac{u_0}{\nu_{bb}+\nu_{bp}+\nu_{bi}} \sim 114.9\ \mathrm{\mu m}. \tag{A1}
\end{equation}
Here $u_0\sim c$ is the beam electron velocity,  $\nu_{bb}=4\sqrt{2\pi}n_{b0}e^4\ln \Lambda_{bb} / 48\pi^2\varepsilon_{0}^2m_e^{1/2}T_b^{3/2}\sim 3.9\times 10^5\ \mathrm{Hz}$ is the beam electron-beam electron collision frequency, $\nu_{bp}=n_pe^4\ln \Lambda_{bp} / 4\pi\varepsilon_{0}^2m_e^2u_0^3 \sim 1.1\times 10^{11}\ \mathrm{Hz}$ is the beam electron-skeleton electron collision frequency, and $\nu_{bi}=(n_{\mathrm{Si}}Z_{\mathrm{Si}}^2+n_{\mathrm{O}}Z_{\mathrm{O}}^2)e^4\ln \Lambda_{bi}/4\pi\varepsilon_{0}^2m_e^2u_0^3 \sim 2.5\times 10^{12} \mathrm{Hz}$ is the beam electron-skeleton ion collision frequency. $T_b \sim 2\ \mathrm{MeV}$ is a characteristic beam temperature due to scatterings from the self-generated fields, $n_p \sim 2.6\times 10^{29}\ \mathrm{m^{-3}}$, $Z_{\mathrm{Si}}=14$ and $Z_{\mathrm{O}}=8$ are the ionization states for Si and O atoms, and $\ln \Lambda_{bb} \sim 22$, $\ln \Lambda_{bp} \sim 13.6$ and $\ln \Lambda_{bi} \sim 11.6$ are the Coulomb logarithms. Note that  here we assume the skeleton atoms are fully ionized, since it gives the largest $\nu_{bp}$ and $\nu_{bi}$ in the system. Likewise, one can obtain the mean free path for the background electrons as $\lambda_{\mathrm{mfp},p}\sim 15.5\ \mathrm{\mu m}$. Since $\lambda_{\mathrm{mfp},b(p)}$ is much larger than the skeleton thickness of 1 $\mu$m, the influence of collisions on REB transport can be neglected. In addition, Eq. \eqref{eq5} suggests that REB branching is independent
of $n_p$. Thus, collisional ionization, of which the main effect here is to increase $n_p$, should also play a negligible role in REB branching. The above arguments have been substantiated with additional simulations \cite{Sup}.

\vspace{1em}

\textit{Appendix B: REB branching versus beam-plasma instabilities.\textemdash }There are several fundamental differences, as listed below, between REB branching and beam-plasma instabilities.

(1) The onset mechanisms are different. REB branching is determined by the uneven fields associated with the target microstructure and can be considered as mainly linear. In particular, no waves (especially beam eigenmodes) are excited. On the other hand, filamentation or modulational instabilities are due to plasma interaction with the beam-excited waves (beam modes) or convective cells, which do not form in the foam plasma because of the pre-existing rigid skeleton-and-pore structure.

(2) The scaling laws of the beam and target parameters are different. In particular, REB branching depends strongly on $l_c$ of the foam-skeleton structure. Furthermore, REB branching (specifically $d_0$) is independent of the beam temperature $T_e$ \cite{Sup}, whereas the growth rate of beam-plasma instabilities decreases with increasing $T_e$ \cite{Bret,Bret1}. In addition, Eq. \eqref{eq5} indicates that REB branching is independent of $n_p$. By contrast, beam-plasma instabilities can be sensitive to it \cite{Bret}.

(3) The density patterns of the beam electrons are different. In REB branching, the lateral separation between two adjacent caustics is almost time-independent and about $l_c$. Downstream the caustics, the random superposition of different branches results in a somewhat homogeneous but fluctuated density pattern. In beam-plasma instabilities, the filaments are separated at kinetic scales at early stages and can then merge into larger ones with an increased separation.

(4) The parameter spaces for the onset of REB branching and beam-plasma instabilities are different. For example, Weibel instability occurs most likely when $n_{b0} \sim n_p$. Resistive filamentation instability requires that collision effects dominate (e.g., in dense foams) and usually occurs at large scales (several-hundred micrometers or larger for hot background). Ionization front instability also occurs at large scales. By contrast, REB branching requires the presence of non-negligible target microstructures (e.g., in low-density porous foam), and it can occur within several tens of micrometers in collisionless systems when $n_{b0} \ll n_p$.


\begin{thebibliography}{10}
\bibitem{note}
Direct laser acceleration can generate high-current relativistic electron beams with parameters comparable to that used in our simulations. See the references \onlinecite{Rosmej,Rosmej1,Huang}.
\bibitem{comment1}
The phase space and energy spectrum of REB in porous medium are significantly broadened, resulting in a rapid increase in the REB temperature, as compared with the homogenous case with similar average density (see the reference \onlinecite{Sup}, Sec. III).
\end{thebibliography}
\end{document}